\documentclass[10pt,a4paper]{article}

\usepackage{amsmath,amssymb}
\usepackage{graphicx}
\usepackage{tabularx}
\usepackage{booktabs}
\usepackage{titling}
\usepackage{url}
\usepackage[left=2.9cm,right=2.9cm,top=2cm,bottom=2cm]{geometry}
\usepackage{color} 
\usepackage{changes}

\colorlet{Changes@Color}{red}

\begin{document}

\author{Martin Roessler \and Jochen Schmitt \and Olaf Schoffer}
\title{Can we trust the standardized mortality ratio? A formal analysis and evaluation based on axiomatic requirements}

\begin{titlingpage}
\maketitle
\begin{abstract}
\noindent {\bf Background}: The standardized mortality ratio (SMR) is often used to assess and compare hospital performance. While it has been recognized that hospitals may differ in their SMRs due to differences in patient composition, there is a lack of rigorous analysis of this and other - largely unrecognized - properties of the SMR.

\noindent {\bf Methods}: This paper proposes five axiomatic requirements for adequate standardized mortality measures: strict monotonicity (monotone relation to actual mortality rates), case-mix insensitivity (independence of patient composition), scale insensitivity (independence of hospital size), equivalence principle (equal rating of hospitals with equal actual mortality rates in all patient groups), and dominance principle (better rating of unambiguously better performing hospitals). Given these axiomatic requirements, effects of variations in patient composition, hospital size, and actual and expected mortality rates on the SMR were examined using basic algebra and calculus. In this regard, we distinguished between standardization using expected mortality rates derived from a different dataset (external standardization) and standardization based on a dataset including the considered hospitals (internal standardization). The results were illustrated by hypothetical examples.

\noindent {\bf Results}: Under external standardization, the SMR fulfills the axiomatic requirements of strict monotonicity and scale insensitivity but violates the requirement of case-mix insensitivity, the equivalence principle, and the dominance principle. All axiomatic requirements not fulfilled under external standardization are
also not fulfilled under internal standardization. In addition, the SMR under internal standardization is scale sensitive and violates the axiomatic requirement of strict monotonicity. 

\noindent {\bf Conclusions}: The SMR fulfills only two (none) out of the five proposed axiomatic requirements under external (internal) standardization. Generally, the SMRs of hospitals are differently affected by variations in case mix and actual and expected mortality rates unless the hospitals are identical in these characteristics. These properties hamper valid assessment and comparison of hospital performance based on the SMR.
\end{abstract}
\end{titlingpage}

\newpage

\section*{Introduction}
Assessing quality of care in hospitals is of high interest to patients, healthcare professionals, and political decision makers. Consequently, there are multiple attempts to characterize and compare hospitals based on quality indicators \cite{AHRQ2020, CMS2020, IQTIG2020, Mansky2015, NWR2020}. The conceptualization of those indicators usually includes some form of risk adjustment. Utilizing statistical methods and measures, risk adjustment aims to facilitate comparison of hospitals with differences in case mix (e.g.\ different shares of high-risk patient groups) that induce outcome differences between the hospitals irrespectively of the true quality of care. Such adjustment is particularly relevant for quality indicators based on in-hospital mortality, which is one of the most frequently considered hospital outcomes. 

A frequently used measure of risk-adjusted mortality is the standardized mortality ratio (SMR) \cite{Amin2019, Berthelot2016, Heijink2008, Jarman2010, Pross2017, Tambeur2019, Taylor2013}. Using indirect standardization, the SMR relates the observed mortality rate of a hospital to its expected mortality rate. The latter is derived by estimating expected mortality rates for predefined strata of patients (i.e.\ patients with similar risk factors characteristics) and aggregating these stratum-specific expected mortality rates according to the hospital's case mix (details on calculation of the SMR are provided below). In this way, the SMR aims to express actual mortality relative to a benchmark that contains information on the hospital's patient composition and, thus, on the distribution of risk factors for in-hospital mortality.

While the SMR is the dominant measure in empirical applications, some studies highlighted problems related to its estimation. SMR values were found to be sensitive to the choice of the estimation method \cite{LacsonJr2001}, readmission rates \cite{Meacock2019}, differences between hospitals with respect to coding quality \cite{Bottle2011}, and correlation between quality of care and risk factors \cite{Roessler2019}. In some cases, changes in the SMR over time were primarily driven by changes in expected rather than actual mortality rates \cite{Mohammed2013}. Moreover, violations of the assumption of identical relationships between mortality and its risk factors across all analyzed hospitals were shown to induce bias in the estimation of the SMR \cite{Mohammed2009}.

In addition to estimation issues, there is evidence for more general methodological problems related to the construction of the SMR. Notably, the SMR was shown to be case-mix sensitive, implying that two hospitals with identical mortality rates in all patient groups may differ in their SMRs due to differences in patient composition \cite{Manktelow2014}. This property is reflected in the results of multiple studies examining the impact of variations in patient composition on the SMRs of hospitals \cite{Manktelow2014, Glance1999, Glance2000, Kahn2007, Pouw2013}. It is noteworthy that all of these studies draw on empirical and/or simulation approaches. While those analyses can provide evidence on basic properties of the SMR, there is a lack of rigorous, formal analysis. Moreover, properties of the SMR other than case-mix sensitivity have rarely been examined.

Against that background, we systematically investigated and evaluated basic properties of the SMR. In a first step, we proposed general properties that characterize adequate measures of standardized mortality. In a second step, we utilized these proposed characteristics to derive a set of axiomatic requirements that should be fulfilled by standardized mortality measures. Formulation of axiomatic requirements for adequate statistical measures is long-established in literature on measurement of income inequality \cite{Cowell2011} and facilitates clear evaluation of the measures' mathematical properties. In a third step, we examined properties of the SMR by drawing on analytical mathematical methods. This approach allowed us to formally investigate the behavior of the SMR given variations in case mix, hospital size, and actual and expected mortality rates. The insights on properties of the SMR were evaluated with respect to the formulated axiomatic requirements for standardized mortality measures. In this way, this paper clarifies and extends the results of previous analyses by providing a comprehensive, systematic, and transparent examination and assessment of the SMR's basic properties. 

\section*{Methods}

All formal analyses relied on basic algebra and differential calculus. In preparation of these analyses, the following sections outline the definition and the analyzed properties of the SMR and the notational conventions used throughout this paper.

\subsection*{Definition and interpretation of the standardized mortality ratio}

We considered $H$ hospitals, indexed by $h=1,\dots,H$. Each patient treated in one of these hospitals belonged to one of $S$ strata, indexed by $s=1,\dots,S$. Each stratum represents a group of patients with the same risk factor characteristics. Let $n_{hs} \in \mathbb{N}$ denote the number of patients belonging to stratum $s$ treated in hospital $h$ and $n_{h} = \sum_{s=1}^{S} n_{hs}$ denote the total number of patients treated in hospital $h$. Note that we also refer to $n_h$ as a measure of hospital size. Furthermore, assume that each hospital was characterized by \textit{actual} stratum-specific mortality rates $p_{hs} \in [0,1]$. Given \textit{expected} stratum-specific mortality rates $p_s^e \in [0,1]$, the SMR of hospital $h$ is defined as the relation between its actual mortality rate $\bar{p}_h = n_h^{-1} \sum_{s=1}^{S} n_{hs} p_{hs}$ and its expected mortality rate $\bar{p}_h^{e} = n_h^{-1} \sum_{s=1}^{S} n_{hs} p_{s}^{e}$, i.e.
\begin{equation}\label{eq:defsmrfirst}
\text{SMR}_h := \frac{\bar{p}_h}{\bar{p}_h^{e}} = \frac{\sum_{s=1}^{S} n_{hs} p_{hs} }{\sum_{s=1}^{S} n_{hs} p_{s}^{e}} .
\end{equation}
Note that while actual mortality rates $p_{hs}$ may be specific for each stratum in a hospital, expected mortality rates $p_s^e$ may only vary by stratum but are the same for all considered hospitals. Hence, the SMR may be interpreted as evaluating actual mortality rates of all hospitals relative to the same ``benchmark'' (i.e. expected) mortality rates, where both actual and expected mortality rates are weighted by each hospital's stratum-specific patient numbers. If the SMR of a hospital exceeds the value of 1, the hospital is judged to perform worse than expected. A SMR smaller than 1 is interpreted as better-than-expected performance. The relative performance of hospitals is often assessed by comparison of their SMRs.

In practice, the hospital-specific mortality rates $p_{hs}$ are unknown and may be estimated by the hospital's \textit{observed} stratum-specific mortality rates. Under this approach, the numerator of Eq~\ref{eq:defsmrfirst} becomes the hospital's observed number of deaths while the denominator is the hospital's expected number of deaths. However, since our analysis does not focus on issues of estimation but examines general properties of the SMR, we treat the actual mortality rates $p_{hs}$ as known (or perfectly estimated) throughout the paper.

\subsection*{Axiomatic requirements for standardized mortality measures}
The objective of this paper is to evaluate properties of the SMR in a systematic way. While Eq~\ref{eq:defsmrfirst} provides the basis for formal analysis, evaluation of the SMR's properties also requires general assumptions on desirable properties of standardized mortality measures. Those properties should be relevant for fair comparison of hospital performance in terms of mortality, which, by assumption, is influenced by the hospitals' care qualities. For this purpose, we propose that a well-behaved measure of standardized mortality should be characterized by the following properties:
\begin{itemize}
\item Increases (decreases) in actual mortality rates should, ceteris paribus, always be reflected in increased (decreased) values of the standardized mortality measure. \\
\textit{Rationale}: Keeping (all relevant) patient-specific risk factors constant, increasing (decreasing) mortality in patients treated in a hospital indicates worse (better) performance of the hospital.
\item The measure should be independent of the hospital's patient composition. \\
\textit{Rationale}: The hospital's case mix does not reflect the hospital's care quality and, thus, should not influence the performance assessment.
\item The measure should be independent of hospital size. \\
\textit{Rationale}: Hospital size per se does not reflect quality of care and, thus, should not influence the performance assessment.
\item The measure should assign the same value to hospitals with identical performance in terms of mortality.  \\
\textit{Rationale}: Fair comparison of hospital performance requires that hospitals with identical care quality may not be evaluated differently.
\item The measure should always rank one hospital better than another hospital if the former unambiguously performs better in terms of care-quality related mortality. \\
\textit{Rationale}: Lower mortality rates of all patient groups in one hospital compared to another hospital imply that each patient's risk of death is lower when being admitted in the former hospital.
\end{itemize}
Based on these necessary properties for valid comparisons of quality of care, we postulate the following five axiomatic requirements for standardized mortality measures:
\begin{itemize}
\item \underline{Strict monotonicity}: Increases (decreases) in a hospital's stratum-specific mortality rates $p_{hs}$ always induce increases (decreases) in the value of the measure assigned to the hospital if the hospital treated patients belonging to that stratum ($n_{hs} > 0$).
\item \underline{Case-mix insensitivity}: Holding actual stratum-specific mortality rates $p_{hs}$ expected mortality rates $p_s^e$ and the hospital's number of patients $n_h$ constant, the value of the measure is insensitive to the hospital's case mix, i.e.\ the hospital's stratum-specific patient shares $n_{hs}/n_h, s=1,\dots,S$.
\item \underline{Scale insensitivity}: Holding case mix ($n_{hs}/n_h, s=1,\dots,S$), actual mortality rates $p_{hs}$, and expected mortality rates $p_s^e$ constant, the measure is insensitive to the hospital's total number of patients $n_h$.
\item \underline{Equivalence principle}: The measure assigns the same value to two hospitals with identical stratum-specific mortality rates $p_{hs}$ or identical deviations of actual stratum-specific mortality rates $p_{hs}$ from expected stratum-specific mortality rates $p_s^e$.
\item \underline{Dominance principle}: The measure always ranks hospital 1 better than hospital 2 if the actual mortality rates of all patient groups treated in hospital 1 are equal to or lower than the mortality rates of these patient groups in hospital 2 ($p_{1s} \leq p_{2s} \, \forall \, s=1,\dots,S$) and the mortality rate of at least one patient group is lower in hospital 1 than in hospital 2 ($\exists  k \in \{1,...,S\} : p_{1k} < p_{2k}$). 
\end{itemize}
Given these axiomatic requirements, we examined effects of variations in case mix, hospital size $n_h$, actual mortality rates $p_{hs}$, and expected mortality rates $p_s^e$ on the SMR. In this regard, it is noteworthy that there are two general ways in which expected mortality rates $p_s^e$ may be derived:
\begin{itemize}
\item \underline{External standardization}: Expected mortality rates may be derived from data that is \textit{not} included in the analysis of the hospitals under consideration, e.g.\ from a dataset of hospitals from a different geographical region. This approach is refereed to as external standardization.
\item \underline{Internal standardization}: Alternatively, expected mortality rates may be derived from the same dataset used to calculate the SMRs of the considered hospitals. In this case, the performance of the hospitals usually is evaluated against their average performance in terms of mortality rates. This approach is referred to as internal standardization. 
\end{itemize}
Taking the difference between external and internal standardization into account, we examined properties of the SMR for both standardization approaches separately. 

\subsection*{Notation}
For notational brevity, arguments of functions are stated explicitly only when they are relevant for the analysis. For instance, the SMR of a specific hospital $h$, which depends on stratum-specific numbers of patients $n_{h1},\dots,n_{hS}$, the hospital's stratum-specific mortality rates $p_{h1},\dots,p_{hS}$, and expected mortality rates $p^e_{1},\dots,p^e_{S}$ is simply written as $\text{SMR}_h$, where
\begin{align}\label{smrshort}
\text{SMR}_h & = \text{SMR}_h(n_{h1},\dots,n_{hS},p_{h1},\dots,p_{hS},p^e_{1},\dots,p^e_{S}) \nonumber \\
& := \frac{\sum_{s=1}^{S} n_{hs} p_{hs} }{\sum_{s=1}^{S} n_{hs} p_{s}^{e}} . 
\end{align}
Adding $\eta \in \mathbb{N}^+$ patients to stratum $s=k$ while holding all other parameters constant is expressed as $\text{SMR}_h(n_{hk}+\eta)$, where
\begin{align}\label{eq:addonesmr}
\text{SMR}_h(n_{hk}+\eta) & \nonumber \\ & = \text{SMR}_h(n_{h1},\dots,n_{h,k-1},n_{hk}+\eta,n_{h,k+1},\dots,n_{hS},p_{h1},\dots,p_{hS},p^e_{1},\dots,p^e_{S}) \nonumber \\
&= \frac{(n_{hk}+\eta)p_{hk} + \sum_{s \neq k} n_{hs} p_{hs} }{(n_{hk}+\eta)p_{k}^e + \sum_{s \neq k} n_{hs} p_{k}^{e}} . 
\end{align}
In the same way, the overall mortality rate of hospital $h$ when multiplying all stratum-specific patient numbers $n_{hs}, s=1,\dots,S$ with a common factor $\lambda \in \mathbb{R}^+$ is written as $\bar{p}_h(\lambda n_{h1},\dots,\lambda n_{hS})$, where
\begin{align}\label{eq:addonepe}
\bar{p}_h(\lambda n_{h1},\dots,\lambda n_{hS}) &= \bar{p}_h(\lambda n_{h1},\dots, \lambda n_{hS},p_{h1},\dots,p_{hS}) \nonumber \\
&= \frac{\sum_{s=1}^{S} (\lambda n_{hs}) p_{hs} }{\sum_{s=1}^{S} (\lambda n_{hs})} . 
\end{align}
To distinguish in notation between external and internal standardization, variables that are affected by the choice of standardization approach are tagged with the superscripts ``ext'' and ``int'', respectively.

\section*{Results}

In the following, effects of variations in case mix, hospital size, and actual and expected mortality rate are examined formally. Analyses were first conducted for the SMR under external standardization and subsequently for the SMR under internal standardization.

\subsection*{External standardization}

As noted above, external standardization refers to the case in which the stratum-specific expected mortality rates are derived from a different dataset. Letting $p_{s}^{e,\text{ext}}$ denote these expected mortality rates, the externally standardized SMR of hospital $h$ is
\begin{equation}\label{eq:defsmr}
\text{SMR}_h^{\text{ext}} := \frac{\bar{p}_h}{\bar{p}_h^{e,\text{ext}}} = \frac{\sum_{s=1}^{S} n_{hs} p_{hs} }{\sum_{s=1}^{S} n_{hs} p_{s}^{e,\text{ext}}} ,
\end{equation}
where $\bar{p}_h^{e,\text{ext}} = n_h^{-1} \sum_{s=1}^{S} n_{hs} p_{s}^{e,\text{ext}}$ is the externally standardized expected mortality rate of hospital~$h$.

\subsubsection*{Variations in case mix under external standardization}

To analyze case-mix sensitivity, we examined effects of a change in a hospital's number of patients belonging to specific strata on Eq~\ref{eq:defsmr} while holding the total number of patients treated in the hospital constant. Formally, we considered a shift of $\eta \in \mathbb{N}^+$ patients from stratum $s=l$, to stratum $s=k$, where $n_{hl} \geq \eta$. The SMR of hospital $h$ thus becomes
\begin{equation}\label{eq:shiftex}
\text{SMR}_h^{\text{ext}}(n_{hk} + \eta, n_{hl} - \eta) =  \frac{(\sum_{s=1}^{S} n_{hs} p_{hs}) + \eta \cdot (p_{hk} - p_{hl})}{(\sum_{s=1}^{S} n_{hs} p_{s}^{e,\text{ext}}) + \eta \cdot (p_k^{e,\text{ext}} - p_l^{e,\text{ext}})} .
\end{equation}
It is noteworthy that Eq~\ref{eq:shiftex} implies that the SMR generally changes due to a shift of patients from stratum $l$ to stratum $k$ even if the hospital's mortality rates in both strata are equal to the expected mortality rates ($p_{hk} = p_k^{e,\text{ext}}, p_{hl} = p_l^{e,\text{ext}})$ as long as $p_{k}^{e,\text{ext}} \neq p_{l}^{e,\text{ext}}$. Hence, performance in line with expected mortality for both strata generally does not imply that the SMR is insensitive to the number of patients belonging to these strata. This result demonstrates the SMR's violation of the axiomatic requirement of case-mix insensitivity. 

For further investigation, the change in the SMR due to the shift in case mix is defined as
\begin{align}
\Omega_{hkl}^{\text{ext}}(\eta) &:= \text{SMR}_h^{\text{ext}}(n_{hk} + \eta, n_{hl} - \eta) - \text{SMR}_h^{\text{ext}}(n_{hk}, n_{hl}) \nonumber \\
&= \frac{(p_{hk} - p_{hl}) - (p_k^{e,\text{ext}} - p_l^{e,\text{ext}}) \cdot \text{SMR}_h^{\text{ext}}(n_{hk}, n_{hl})}{\eta^{-1} n_h \bar{p}_h^{e,\text{ext}}(n_{hk},n_{hl}) + (p_k^{e,\text{ext}} - p_l^{e,\text{ext}}) } . \label{eq:chshiftex}
\end{align}
Eq~\ref{eq:chshiftex} shows that the change in the SMR due to a shift of patients from stratum $l$ to stratum $k$ is, in absolute terms, large if the number of shifted patients $\eta$ is large, the number of patients treated in the hospital $n_h$ is small, and the hospital's overall expected mortality rate $\bar{p}_h^{e,\text{ext}}(n_{hk},n_{hl})$ is low. Since $\bar{p}_h^{e,\text{ext}}(n_{hk},n_{hl})$ depends on the stratum-specific patient numbers $n_{hs}, s=1,\dots,S$, the latter implies that the change in the SMR due to a variation in case mix depends on the initial case mix of the hospital. 

The sign of Eq~\ref{eq:chshiftex} is determined according to
\begin{align}\label{eq:dirchshiftex}
\Omega_{hkl}^{\text{ext}}(\eta) > 0 \quad \text{if} \quad (p_{hk} - p_{hl}) > (p_k^{e,\text{ext}} - p_l^{e,\text{ext}}) \cdot \text{SMR}_h^{\text{ext}}(n_{hk}, n_{hl}) , \\
\Omega_{hkl}^{\text{ext}}(\eta) = 0 \quad \text{if} \quad (p_{hk} - p_{hl}) = (p_k^{e,\text{ext}} - p_l^{e,\text{ext}}) \cdot \text{SMR}_h^{\text{ext}}(n_{hk}, n_{hl}) , \\
\Omega_{hkl}^{\text{ext}}(\eta) < 0 \quad \text{if} \quad (p_{hk} - p_{hl}) < (p_k^{e,\text{ext}} - p_l^{e,\text{ext}}) \cdot \text{SMR}_h^{\text{ext}}(n_{hk}, n_{hl}) . \label{eq:dirchshiftexend}
\end{align}
Hence, the direction of the change in the SMR due to a shift of patients from stratum $l$ to stratum $k$ depends on the difference between the hospital's mortality rates of these strata ($p_{hk} - p_{hl}$), the difference between the strata's expected mortality rates ($p_k^{e,\text{ext}} - p_l^{e,\text{ext}}$) and the hospital's SMR. If the hospital's mortality rate in stratum $k$ is higher than in stratum $l$ ($p_{hk} - p_{hl} > 0$) while the opposite is true for the expected mortality rates ($p_k^{e,\text{ext}} - p_l^{e,\text{ext}} < 0$), the hospital's SMR increases due to the shift of patients, and vice versa. However, if both actual and expected mortality rate differences are positive ($p_{hk} - p_{hl} > 0$ and $p_k^{e,\text{ext}} - p_l^{e,\text{ext}} > 0$), hospitals with high SMRs experience a reduction in the SMR whereas hospitals with low SMRs experience an increase in the SMR. Similarly, concordant negative hospital-specific and expected mortality rate differences ($p_{hk} - p_{hl} < 0$ and $p_k^{e,\text{ext}} - p_l^{e,\text{ext}} < 0$) imply that the SMR of a hospital increases (decreases) if the initial SMR of the hospital is high (low). The SMR generally changes in accordance with the relation between actual and expected mortality rate differences only if $\text{SMR}_h^{\text{ext}}(n_{hk}, n_{hl}) = 1$, as this implies that $\Omega_{hkl}^{\text{ext}}(\eta) \gtreqless 0$ if $(p_{hk} - p_{hl}) \gtreqqless (p_k^{e,\text{ext}} - p_l^{e,\text{ext}})$. 

As noted above, performance in line with expected mortality rates ($p_{hk}=p_k^{e,\text{ext}}$ and $p_{hl}=p_l^{e,\text{ext}}$) does not imply that the SMR is insensitive to the number of patients belonging to the considered strata. Under this condition, $p_{hk} - p_{hl} > 0$ implies that $\Omega_{hkl}^{\text{ext}}(\eta) \gtreqless 0$ if $\text{SMR}_h^{\text{ext}}(n_{hk}, n_{hl}) \lesseqgtr 1$. Thus, a shift of patients from a stratum with a lower to a stratum with a higher mortality rate leads to an increase (decrease) in the SMR if the hospital's SMR initially is smaller (greater) than 1. By the same token, a shift of patients from a stratum with a higher to a stratum with a lower mortality rate $p_{hk} - p_{hl} < 0$ implies that $\Omega_{hkl}^{\text{ext}}(\eta) \gtreqless 0$ if $\text{SMR}_h^{\text{ext}}(n_{hk}, n_{hl}) \gtreqless 1$ if the hospital performs in line with expected mortality rates. In this scenario, the SMR of the hospital increases (decreases) if its initial SMR is above (below) unity.

In the extreme case, in which all patients are concentrated in a specific stratum $k$ ($n_{hk} = n_h$), the hospital's SMR equals the relation between the actual and the observed mortality rate of that stratum, i.e.
\begin{equation}\label{eq:intconent}
\text{SMR}_h^{\text{ext}}(n_{h1}=0,\dots,n_{h,k-1}=0,n_{hk}=n_h,n_{h,k+1}=0,\dots,n_{hS}=0) = \frac{p_{hk}}{p_k^{e,\text{ext}}} .
\end{equation}
For illustration of case-mix sensitivity under external standardization, we considered two hospitals and three strata of patients (Table~\ref{tab:casemixext}). Both hospitals had the same case mix, with $20-\eta$ patients belonging to stratum 1, $\eta$ patients belonging to stratum 2 and 5 patients belonging to stratum 3. The parameter $\eta$ is used to determine the allocation of patients to stratum 1 and stratum 2. If $\eta=0$, both hospitals had 20 patients in stratum 1 and 0 patients in stratum 2. If $\eta = 20$, 20 patients were allocated to stratum 2 while the hospitals have no patient in stratum 1. Furthermore, both hospitals performed in line with expected mortality rates in strata 1 and 2. The only difference between the hospitals is that hospital 1 had a higher-than-expected mortality rate in stratum 3 ($0.2 > 0.15$) while hospital 2 performed better than expected in this stratum ($0.1 < 0.15$). Accordingly, the SMR of hospital 1 exceeds the value of 1 while the SMR of hospital 2 is below unity.

\begin{figure}[h!] \centering
	\includegraphics[width=.8\textwidth]{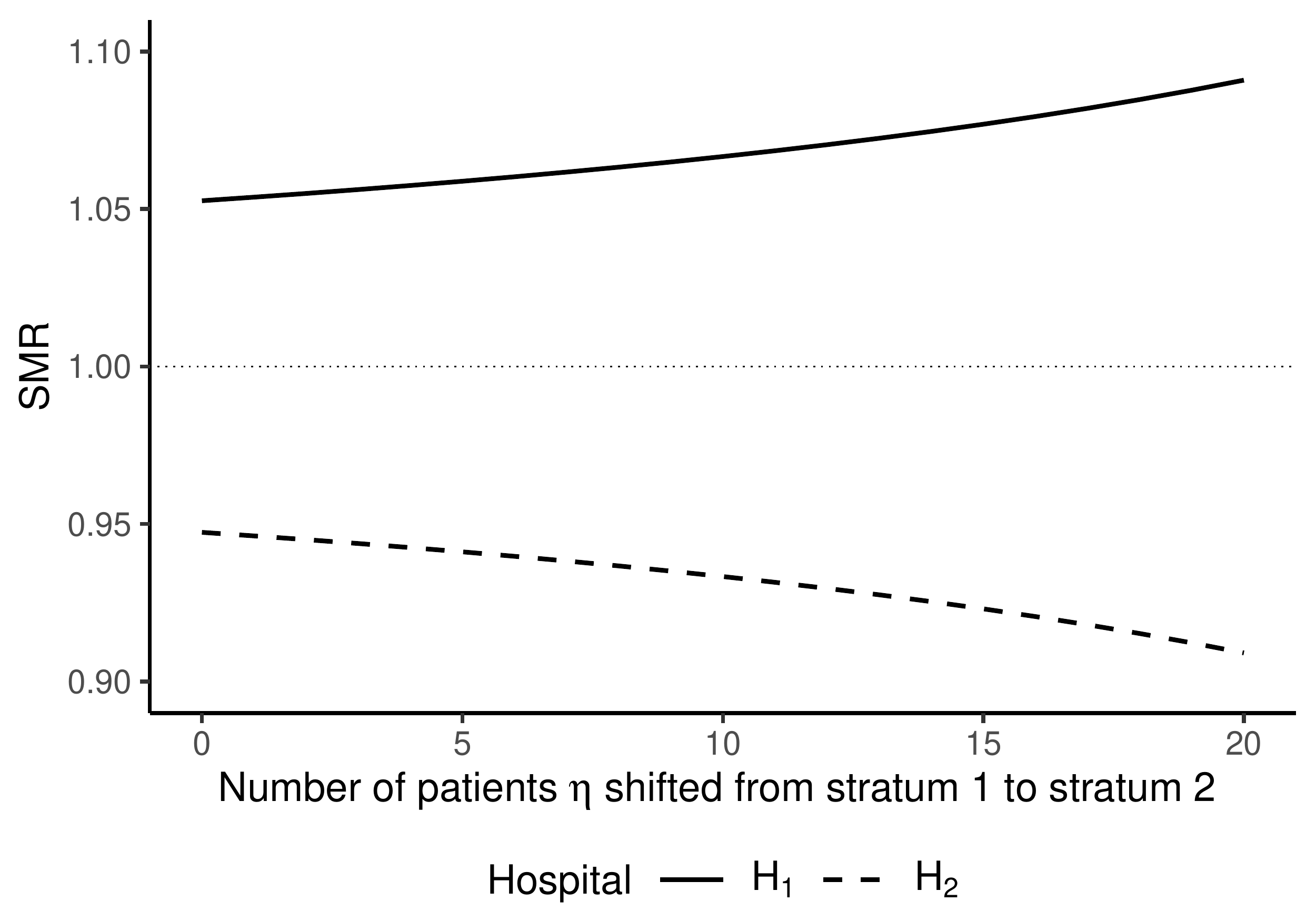}
	\caption{SMRs for different numbers of patients $\eta$ shifted from stratum 1 to stratum 2 in hospital 1, holding the number of patients belonging to stratum 3 constant \label{fig:casemixext}}
\end{figure}

Fig~\ref{fig:casemixext} shows the SMRs of the hospitals for different allocations of patients to strata 1 and 2 as induced by different values of $\eta$. Although both hospitals were identical in case mix and performed in line with expected mortality rates in both affected strata, their SMRs are affected by a shift of patients from stratum 1 to stratum 2. As indicated by Eqs~\ref{eq:dirchshiftex}-\ref{eq:dirchshiftexend}, hospital 1 experiences an increase in its SMR whereas the SMR of hospital 2 decreases when the number of patients allocated to stratum 2 is increased (i.e.\ $\eta$ is increased). This is because mortality in stratum 2 was lower than mortality in stratum 1 and the SMR of hospital 1 exceeds unity while the SMR of hospital 2 is below unity.

\begin{table}[h!]
	\centering
	\caption{Example of variations in case mix under external standardization: parameter values \label{tab:casemixext}}
	\begin{tabular}{c|cc|cc|c} \toprule
		Stratum & \multicolumn{2}{c}{Hospital 1 ($H_1$)} & \multicolumn{2}{c}{Hospital 2 ($H_2$)} & Exp. mortality rate \\ \midrule
		s     & $n_{1s}$ & $p_{1s}$ & $n_{2s}$ & $p_{2s}$ & $p_s^{e,\text{ext}}$ \\ \midrule
		1     & 20$-\eta$ & 0.2   & 20$-\eta$ & 0.2   & 0.2 \\
		2     & $\eta$ & 0.1   & $\eta$ & 0.1   & 0.1 \\
		3     & 5     & 0.2   & 5     & 0.1  & 0.15 \\ \bottomrule
	\end{tabular}%
\end{table}%

\subsubsection*{Variations in hospital size under external standardization}

To examine variations in hospital size, we considered a proportional shift in the numbers of patients treated in all strata of a hospital by the scale factor $\lambda \in \mathbb{R}^+$, where $\lambda = 1$ is the initial scale of the hospital. For $\lambda > 1$, this reflects a situation in which the total number of patients treated in the hospital is increased by factor $\lambda$ while the case mix (i.e.\ the shares of the strata in the hospital's total number patients) is held constant. For the SMR under external standardization is follows that
\begin{align}\label{eq:scaleex}
\text{SMR}_h^\text{ext}(\lambda n_{h1},\dots,\lambda n_{hS}) &= \frac{\sum_{s=1}^{S} \lambda n_{hs} p_{hs}}{\sum_{s=1}^{S} \lambda n_{hs} p_s^{e,\text{ext}}} \nonumber \\
&= \frac{\sum_{s=1}^{S} n_{hs} p_{hs}}{\sum_{s=1}^{S} n_{hs} p_s^{e,\text{ext}}} = \text{SMR}_h^\text{ext}(n_{h1},\dots,n_{hS}) .
\end{align} 
Since the value of the scaled SMR is the same as the value of the original SMR, the SMR fulfills the axiomatic requirement of scale insensitivity under external standardization. Increases in hospital size do not change the value of the SMR, ceteris paribus.

Scale insensitivity under external standardization is illustrated by the example of a hospital with two strata, containing 20 and 40 patients, respectively, in the initial situation (Table~\ref{tab:scaleext}). The mortality rates were assumed to be 0.05 in the first and 0.15 in the second stratum. Expected mortality rates in both strata were fixed at the value of 0.1. In the initial situation, this corresponds to 7 observed and 6 expected deaths, which results in a SMR of 1.17. Doubling the size of the hospital while holding case mix constant ($\lambda = 2$) doubles both, the number of patients and the number of deaths in each stratum. However, the SMR of the hospital remains constant at the value of 1.17. The same is true for further increases of hospital size as induced by higher values of $\lambda$.

\begin{table}[htbp]
	\centering
\caption{Example of variations in hospital size under external standardization \label{tab:scaleext}}
	\begin{tabular}{lrrrrr}
		\toprule
		Quantity & \multicolumn{1}{l}{initial} & \multicolumn{1}{l}{$\lambda = 2$} & \multicolumn{1}{l}{$\lambda = 3$} & \multicolumn{1}{l}{$\lambda = 4$} & \multicolumn{1}{l}{$\lambda = 5$} \\
		\midrule
		Patients in stratum 1 $n_{h1}$ & 20    & 40    & 60    & 80    & 100 \\
		Patients in stratum 2 $n_{h2}$ & 40    & 80    & 120   & 160   & 200 \\
		\midrule
		Actual number of deaths $\sum_{s=1}^S n_{hs} p_{hs}$ & 7     & 14    & 21    & 28    & 35 \\
		Exp. number of deaths $\sum_{s=1}^S n_{hs} p_{s}^{e,\text{ext}}$ & 6     & 12    & 18    & 24    & 30 \\
		\midrule
		$\text{SMR}_h^{\text{ext}}$ & 1.17  & 1.17  & 1.17  & 1.17  & 1.17 \\
		\bottomrule
	\end{tabular}%

$p_{h1} = 0.05, p_{h2} = 0.15, p_{1}^{e,\text{ext}} =  p_{2}^{e,\text{ext}} = 0.1$
\end{table}%

\subsubsection*{Variations in actual mortality rates under external standardization}

Effects of variations in actual mortality rates were examined by calculating the marginal effect (i.e.\ the partial derivative) \cite{Jones2007} of an increase in the mortality rate of stratum $k$ in hospital $h$ on the hospital's SMR:
\begin{equation}\label{eq:me_ext}
\text{ME}_{h,p_{hk}}^{\text{ext}} := \frac{\partial \text{SMR}_h^{\text{ext}}}{\partial p_{hk}} = \frac{n_{hk}}{n_h} \cdot \frac{1}{\bar{p}_h^{e,\text{ext}}} .
\end{equation}
If $n_{hk} > 0$, Eq~\ref{eq:me_ext} implies that $\text{ME}_{h,p_{hk}}^{\text{ext}} > 0$, i.e. an increase in the mortality rate of a specific stratum increases the SMR of the hospital. The SMR under external standardization therefore fulfills the axiomatic requirement of strict monotonicity. The increase in the SMR induced by an increase in the stratum-specific mortality rate is relatively large (small) if the patients included in the stratum account for a large (small) share $n_{hk}/n_h$ of patients treated in the hospital. Furthermore, the marginal effect decreases in the hospital's expected overall mortality rate $\bar{p}_h^{e,\text{ext}}$. The latter implies that an increase in stratum-specific mortality generally affects hospitals differently, as $\bar{p}_h^{e,\text{ext}}$ depends on a hospital's case mix.

This result also applies in the case in which the hospital's actual mortality rates of all strata are increased by the absolute amount of $\mathrm{d}p$. This corresponds to a situation in which the overall mortality rate of the hospital is increased by $\mathrm{d}p$. Calculating the differential of Eq~\ref{eq:defsmr} in all actual mortality rates and using $\mathrm{d}p_{hs} = \mathrm{d}p, s=1,\dots,S$ yields
\begin{equation}\label{eq:totaldiffext}
\mathrm{d} \text{SMR}_h^{\text{ext}} = \sum_{s=1}^{S} \frac{\partial \text{SMR}_h^{\text{ext}}}{\partial p_{hs}} \mathrm{d}p_{hs} = \frac{\mathrm{d}p}{\bar{p}_h^{e,\text{ext}}} .
\end{equation}
Similar to an increase in the mortality rate of a single stratum, increases in the mortality rates of all strata have a large (small) impact on the hospital's SMR when the hospital's overall expected mortality rate is small (large).

The results on mortality rate variations under external standardization are illustrated by the example of two hospitals and three strata of patients (Table~\ref{tab:actmortext}). Both hospitals treated 10 patients, with 5 belonging to stratum 1. The difference between the hospitals was that the remaining 5 patients of hospital 1 belonged to stratum 2 whereas those of hospital 2 belonged to stratum 3. In all strata, the hospitals performed in line with expected mortality rates, such that the SMR of both hospitals in the initial situation is 1.

\begin{figure}[h!] \centering
	\includegraphics[width=.8\textwidth]{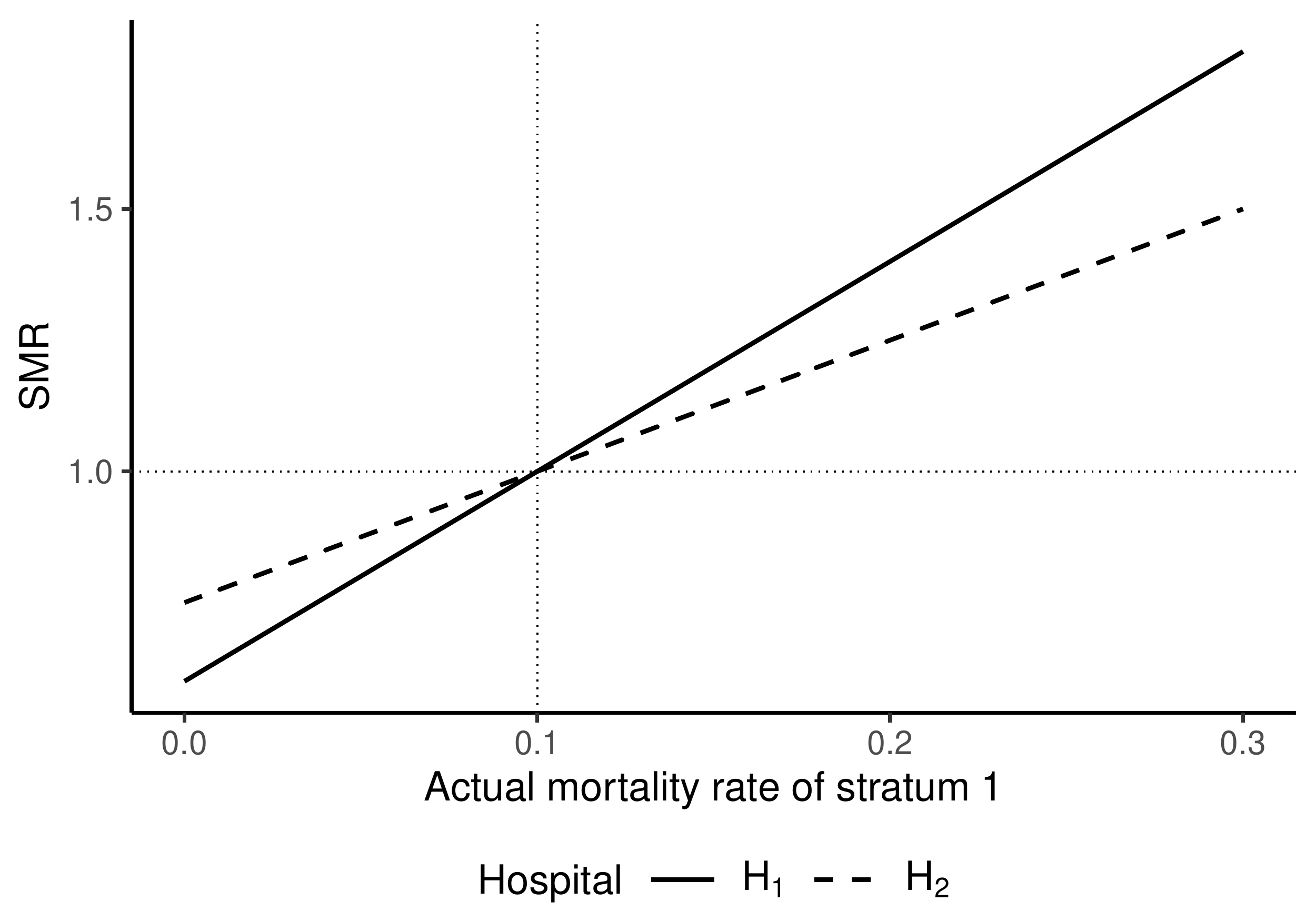}
	\caption{SMRs for different mortality rates $p_{11} = p_{21}$  \label{fig:mort_ext}}
\end{figure}

Holding the remaining parameter values constant, Fig~\ref{fig:mort_ext} shows the SMRs of the hospitals for different mortality rates in stratum 1. Note that the mortality rates in stratum 1 were varied simultaneously for hospital 1 and hospital 2 in each scenario ($p_{11} = p_{21}$), such that there is no difference in the performance of the hospitals with respect to stratum 1. While the SMRs of both hospitals are equal in the initial situation, lower-than-expected mortality rates in stratum 1 ($p_{11} = p_{21} < 0.1$) imply that the SMR of hospital 1 is lower than the SMR of hospital 2. For higher-than-expected mortality rates ($p_{11} = p_{21} > 0.1$), the SMR of hospital 1 is higher than the SMR of hospital 2. The reason for this result is that (actual and expected) mortality rates of hospital 2 in stratum 3 are higher than (actual and expected) morality rates of hospital 1 in stratum 2. This implies that the expected overall mortality rate of hospital 2 is higher than the expected overall mortality rate of hospital 1. According to Eq~\ref{eq:me_ext}, this implies that the SMR of hospital 1 reacts more sensitive to changes in case mix than the SMR of hospital 2. The example therefore demonstrates that two hospitals with identical deviations of actual from expected mortality rates generally do not have the same SMR value. Hence, the SMR under external standardization violates the equivalence principle.

\begin{table}[h!]
	\centering
	\caption{Example of variations in actual mortality rates under external standardization: initial parameter values}
	\begin{tabular}{c|cc|cc|c}
		\toprule
		 Stratum & \multicolumn{2}{c|}{Hospital 1 ($H_1$)} & \multicolumn{2}{c|}{Hospital 2 ($H_2$)} & Exp. mortality rate \\ \midrule
		$s$ & $n_{1s}$   & $p_{1s}$   & $n_{2s}$   & $p_{2s}$   & $p_s^\text{ext}$ \\
		\midrule
		$1$ & 5     & $p_{11}=$ 0.1   & 5     & $p_{21}=$ 0.1   & 0.1 \\
		$2$ & 5     & 0.15   & -     & -     & 0.15 \\
		$3$ & -     & -     & 5     & 0.3   & 0.3 \\
		\bottomrule
	\end{tabular}%
	\label{tab:actmortext}%
\end{table}%

\subsubsection*{Variations in expected mortality rates under external standardization}

Effects of variations in expected mortality rates on the SMR under external standardization were revealed by calculating the marginal effect of an increase in the stratum-specific expected mortality rate $p_k^{e,\text{ext}}$:
\begin{equation}\label{eq:margexpext}
\text{ME}_{h,p_{k}^e}^{\text{ext}} := \frac{\partial \text{SMR}_h^{\text{ext}}}{\partial p_{k}^{e,\text{ext}}} = - \text{SMR}_h^{\text{ext}} \cdot \frac{n_{hk}}{n_h} \cdot \frac{1}{\bar{p}_h^{e,\text{ext}}} .
\end{equation}
According to Eq~\ref{eq:margexpext}, $n_{hk}>0$ implies that $\text{ME}_{h,p_{k}^e}^{\text{ext}} < 0$, i.e. an increase in the expected mortality rate of a stratum reduces the hospital's SMR if it treated patients belonging to this stratum. This reduction is (in absolute terms) larger for hospitals with higher SMRs, a larger share $n_{hk}/n_k$ of patients belonging the considered stratum, and lower expected overall mortality rates $\bar{p}_h^{e,\text{ext}}$. Thus, effects of variations in stratum-specific expected mortality rates depend on the hospital's case mix and the initial value of the hospital's SMR.

The same applies to an increase in all stratum-specific expected mortality rates by the absolute amount of $\mathrm{d}p_s^{e,\text{ext}} = \mathrm{d}p, s=1,\dots,S$ as the associated change in the SMR depends on both the size of the hospital's SMR and the expected overall mortality rate:
\begin{equation}\label{eq:allexpmortext}
\mathrm{d} \text{SMR}_h^{\text{ext}} = \sum_{s=1}^{S}  \frac{\partial \text{SMR}_h^{\text{ext}}}{\partial p_{s}^{e,\text{ext}}} \mathrm{d}p_s^{e,\text{ext}} = - \text{SMR}_h^{\text{ext}} \cdot \frac{\mathrm{d}p}{\bar{p}_h^{e,\text{ext}}} .
\end{equation}
To illustrate the effects of changes in expected mortality rates under external standardization, we considered two hospitals and two strata of patients (Table~\ref{tab:exexpext}). Both hospitals had 5 patients with a mortality rate of 0.1 in stratum 1. The hospitals differed with respect to stratum 2, where hospital hospital 1 had 5 patients with a mortality rate of 0.2 and hospital hospital 2 had 15 patients with a mortality rate of 0.15. Note that both hospitals performed worse than expected in stratum 2 as the expected mortality rate was 0.1. Overall, hospital 2 was performing better than hospital 1 due to equal actual mortality rates in stratum 1 and a lower mortality rate in stratum 2.

\begin{figure}[h!] \centering
	\includegraphics[width=.8\textwidth]{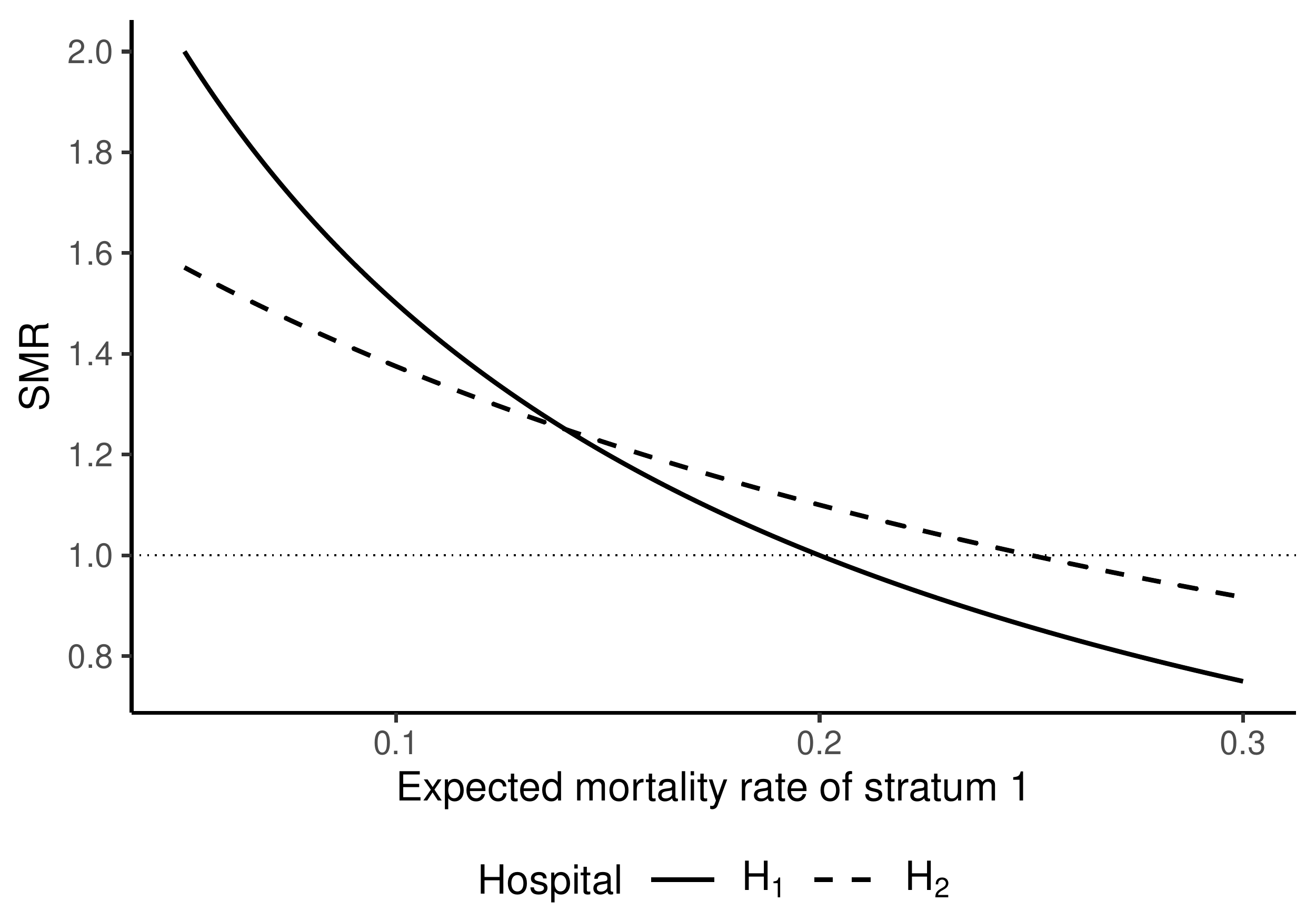}
	\caption{SMRs for different expected mortality rates $p_1^{e,\text{ext}}$  \label{fig:expmort_ext}}
\end{figure}

Fig~\ref{fig:expmort_ext} shows the effect of varying the expected mortality rate of stratum 1 $p_1^{e,\text{ext}}$ on the SMRs of the hospitals. Starting at low expected mortality rates of stratum 1, the SMR of hospital 1 is higher than the SMR of hospital 2, implicating that hospital 2 performed better than hospital 1. Increasing the expected mortality rate of stratum 1 reduces the SMRs of both hospitals. However, since hospital 1 has a higher share of patients in stratum 1 and a higher initial SMR, it experiences a stronger decrease in its SMR. At an expected mortality rate of $p_1^{e,\text{ext}}=0.14$, the SMRs of both hospitals are equal. For further increased expected mortality rates of stratum 1, the SMR of hospital 1 becomes lower than the SMR of hospital 2 although the overall performance of hospital 2 was better than the performance of hospital 1. This result is driven by the fact that stratum 1 accounts for a higher share of patients in hospital 1 than in hospital 2. By the virtue of Eq~\ref{eq:margexpext}, this implies that hospital 1 ``benefits'' more from increases in the expected mortality rate of this stratum in terms of reductions in the SMR even if the SMRs of both hospitals are equal. With respect to the formulated axiomatic requirements, the example therefore demonstrates that the SMR under external standardization violates the dominance principle.

\begin{table}[h!]
	\centering
	\caption{Example of variations in expected mortality rate under external standardization: parameter values}
	\begin{tabular}{c|cc|cc|c}
		\toprule
		Stratum & \multicolumn{2}{c|}{Hospital 1 ($H_1$)} & \multicolumn{2}{c|}{Hospital 2 ($H_2$)} &  Exp. mortality \\ \midrule
		$s$ & $n_{1s}$   & $p_{1s}$   & $n_{2s}$   & $p_{2s}$   & $p_s^{e,\text{ext}}$ \\
		\midrule
		$1$ & 5     & 0.1   & 5     & 0.1   & $p_1^{e,\text{ext}}$ \\
		$2$ & 5     & 0.2   & 15     & 0.15     & 0.1 \\
		\bottomrule
	\end{tabular}%
	\label{tab:exexpext}%
\end{table}%

\section*{Internal standardization}

The stratum-specific mortality rates $p_s^{e,\text{int}}$ were calculated from the same dataset used for later analysis when derived by internal standardization. The SMR of hospital $h$ based on the internal standard therefore is expressed as
\begin{equation}\label{eq:smr_int}
\text{SMR}_h^{\text{int}} = \frac{\bar{p}_h}{\bar{p}_h^{e,\text{int}}} = \frac{\sum_{s=1}^{S} n_{hs} p_{hs} }{\sum_{s=1}^{S} n_{hs}p_s^{e,\text{int}}}, 
\end{equation}
where $\bar{p}_h^{e,\text{int}} = n_h^{-1} \sum_{s=1}^{S} n_{hs}p_s^{e,\text{int}}$ is the hospital's internally standardized expected mortality rate. Since expected mortality rates were derived from the same dataset used for calculation of the hospitals' SMRs, they implicitly depend on the stratum-specific mortality rates $p_{js}$ and the number of patients $n_{js}$ of the included hospitals $j=1,\dots,H$. The internal standard may be chosen such that the expected mortality rate of each stratum equals the weighted average mortality rate of that stratum across all hospitals, i.e.
\begin{equation}\label{eq:avmort}
p_s^{e,\text{int}} := \bar{p}_s = \frac{1}{n_s} \sum_{j=1}^{H} n_{js} p_{js} ,
\end{equation}
where $n_s = \sum_{j=1}^{H} n_{js}$ is the total number of patients in stratum $s$. In terms of interpretability, this approach to standardization has the advantage that a hospital with average mortality rates in all strata ($p_{hs} = \bar{p}_s, s=1,\dots,S$) has a $\text{SMR}_h^{\text{int}} = 1$.

\subsubsection*{Variations in case mix under internal standardization}

Given a shift of $\eta$ patients from stratum $l$ to stratum $k$, the change in the SMR under internal standardization is defined as
\begin{align}
\Omega_{hkl}^{\text{int}}(\eta) &:= \text{SMR}_h^{\text{int}}(n_{hk} + \eta, n_{hl} - \eta) - \text{SMR}_h^{\text{int}}(n_{hk}, n_{hl}) \nonumber \\
&= \frac{(p_{hk} - p_{hl}) - (\tilde{p}_{hk} - \tilde{p}_{hl}) \cdot \text{SMR}_h^{\text{int}}(n_{hk}, n_{hl})}{\eta^{-1} n_h \bar{p}_h^{e,\text{int}}(n_{hk},n_{hl}) + (\tilde{p}_{hk} - \tilde{p}_{hl}) } . \label{diffsmrint} 
\end{align}
This expression analogous to Eq~\ref{eq:chshiftex} with the exogenous expected mortality rates $p_k^{e,\text{ext}}, p_l^{e,\text{ext}}$ replaced by
\begin{align}\label{eq:thresratesint}
\tilde{p}_{hk} &= \alpha_{hk} p_{hk} + (1-\alpha_{hk}) \bar{p}_k(n_{hk}) , \\
\tilde{p}_{hl} &= \alpha_{hl}  p_{hl} + (1-\alpha_{hl}) \bar{p}_l(n_{hl}) .
\end{align}
Hence, $\tilde{p}_{hk}$ and $\tilde{p}_{hl}$ represent weighted averages of the hospital's stratum-specific mortality rates ($p_{hk}$, $p_{hl}$) and the respective average stratum-specific mortality rates ($\bar{p}_k$, $\bar{p}_l$). The weights $\alpha_{hk} = (n_{hk} + \eta)/(n_k + \eta)$ and $\alpha_{hl} = (n_{hl} - \eta)/(n_l - \eta)$ reflect the degree to which hospital $h$ accounts for the total number of patients in the considered strata. Similar to the SMR under external standardization, the SMR under internal standardization generally changes due to a change in case mix. Thus, it does not fulfill the axiomatic requirement of case-mix insensitivity.

From Eq~\ref{diffsmrint} follows that
\begin{align}\label{eq:mechint}
\Omega_{hkl}^{\text{int}}(\eta) > 0 \quad \text{if} \quad  (p_{hk}-p_{hl}) > (\tilde{p}_{hk} - \tilde{p}_{hl}) \cdot \text{SMR}_h^{\text{int}}(n_{hk}, n_{hl}) ,\\
\Omega_{hkl}^{\text{int}}(\eta) = 0 \quad \text{if} \quad (p_{hk}-p_{hl}) =  (\tilde{p}_{hk} - \tilde{p}_{hl}) \cdot \text{SMR}_h^{\text{int}}(n_{hk}, n_{hl}) , \label{eq:mechintmid} \\
\Omega_{hkl}^{\text{int}}(\eta) < 0 \quad \text{if} \quad (p_{hk}-p_{hl}) <  (\tilde{p}_{hk} - \tilde{p}_{hl}) \cdot \text{SMR}_h^{\text{int}}(n_{hk}, n_{hl}) . \label{eq:mechintend}
\end{align}
Similar to the results for case-mix variations under external standardization (Eqs~\ref{eq:dirchshiftex}-\ref{eq:dirchshiftexend}), the direction of change in the SMR induced by a shift of patients from stratum $l$ to stratum $k$ depends on the difference of the actual stratum-specific mortality rates ($p_{hk}-p_{hl}$) and the SMR-weighted difference in the endogenous threshold mortality rates ($\tilde{p}_{hk} - \tilde{p}_{hl}$). 

In the extreme case in which the hospital accounts for the total number of patients in both strata ($n_{hk} = n_k, n_{hl} = n_l)$, it holds that $\alpha_{hk} = \alpha_{hl} = 1$, which implies that $\tilde{p}_{hk} = p_{hk}$ and $\tilde{p}_{hl} = p_{hl}$. For $\text{SMR}_h^{\text{int}}(n_{hk}, n_{hl}) > 0$ follows that
\begin{align}\label{eq:casealpha1}
\Omega_{hkl}^{\text{int}}(\eta) \gtrless 0 \quad  &\text{if} \quad p_{hk}-p_{hl} > 0 \quad \text{and} \quad \text{SMR}_h^{\text{int}}(n_{hk}, n_{hl}) \lessgtr 1 , \\
\Omega_{hkl}^{\text{int}}(\eta) = 0 \quad &\text{if} \quad p_{hk}-p_{hl} = 0, \\
\Omega_{hkl}^{\text{int}}(\eta) \gtrless 0 \quad &\text{if} \quad  p_{hk}-p_{hl} < 0  \quad \text{and} \quad \text{SMR}_h^{\text{int}}(n_{hk}, n_{hl}) \gtrless 1 . \label{eq:casealpha1end}
\end{align}
Hence, a shift of patients from a stratum with a lower to a stratum with a higher mortality rate increases (decreases) the SMR of hospitals with below-average (above-average) SMRs. Similarly, a shift of patients from a stratum with a higher to a stratum with a lower mortality rate decreases (increases) the SMR of hospitals with below-average (above-average) SMRs. These results are driven by assumption that the hospital fully serves as its own reference in both strata. Hence, a concentration of patients in a stratum with a relatively high actual mortality rate implies a greater ``benefit'' in terms of a lower SMR for hospitals with above-average SMRs and vice versa.

In the other extreme case, the hospital accounts for a negligible share of the strata's total number of patients. Holding $n_{hk}$ and $n_{hl}$ constant, it can be derived that $\lim_{n_k \rightarrow \infty} \alpha_{hk} = \lim_{n_k \rightarrow \infty} \alpha_{hl}= 0$, which implies that  $\lim_{n_l \rightarrow \infty} \tilde{p}_{hk} = \bar{p}_k(n_{hk})$ and $\lim_{n_l \rightarrow \infty} \tilde{p}_{hl} = \bar{p}_l(n_{hl})$. Thus, $\text{SMR}_h^{\text{int}}(n_{hk}, n_{hl}) > 0$ implies asymptotically that
\begin{align}\label{eq:casealpha0}
\Omega_{hkl}^{\text{int}}(\eta) \gtrless 0 \quad  &\text{if} \quad \bar{p}_k(n_{hk})-\bar{p}_l(n_{hl}) > 0 \quad \text{and} \quad \text{SMR}_h^{\text{int}}(n_{hk}, n_{hl}) \lessgtr 1 , \\
\Omega_{hkl}^{\text{int}}(\eta) = 0 \quad &\text{if} \quad \bar{p}_k(n_{hk})-\bar{p}_l(n_{hl}) = 0, \\
\Omega_{hkl}^{\text{int}}(\eta) \gtrless 0 \quad  &\text{if} \quad \bar{p}_k(n_{hk})-\bar{p}_l(n_{hl}) < 0 \quad \text{and} \quad \text{SMR}_h^{\text{int}}(n_{hk}, n_{hl}) \gtrless 1 , \label{eq:casealpha0end}
\end{align}
For large values of $n_k$ and $n_h$ relative to $n_{hk}$ and $n_{hl}$, respectively, the stratum-specific average mortality rates $\bar{p}_k(n_{hk})$ and $\bar{p}_l(n_{hl})$ are almost exclusively determined by the mortality rates of hospitals other than $h$. Hence, the behavior of the SMR under internal standardization is similar to the behavior of the SMR under external standardization if the hospital accounts for negligible shares of the strata's total numbers of patients because the hospital has little influence in the internal standard.

For illustration, we considered the case of two hospitals and two strata of patients (Table~\ref{tab:casemixint}). With regard to stratum 1, both hospitals were characterized by a mortality rate of 0.1, implying that the expected mortality rate of stratum 1 is also 0.1. With respect to stratum 2, hospital 1 was characterized by a higher mortality rate than hospital 2 ($0.3 > 0.1$). While the patient numbers of hospital 2 allocated to the strata were fixed at values of 25 and 10, respectively, the parameter $\eta$ determined the number of patients treated in hospital 1 belonging to stratum 1 and 2, respectively. If $\eta=0$, all patients of hospital 1 were allocated to stratum 1 and no patient was allocated to stratum 2. If $\eta=50$, no patient of stratum 1 was treated in hospital 2 while the hospital treated 50 patients belonging to stratum 2.

\begin{figure}[h!] \centering
	\includegraphics[width=.8\textwidth]{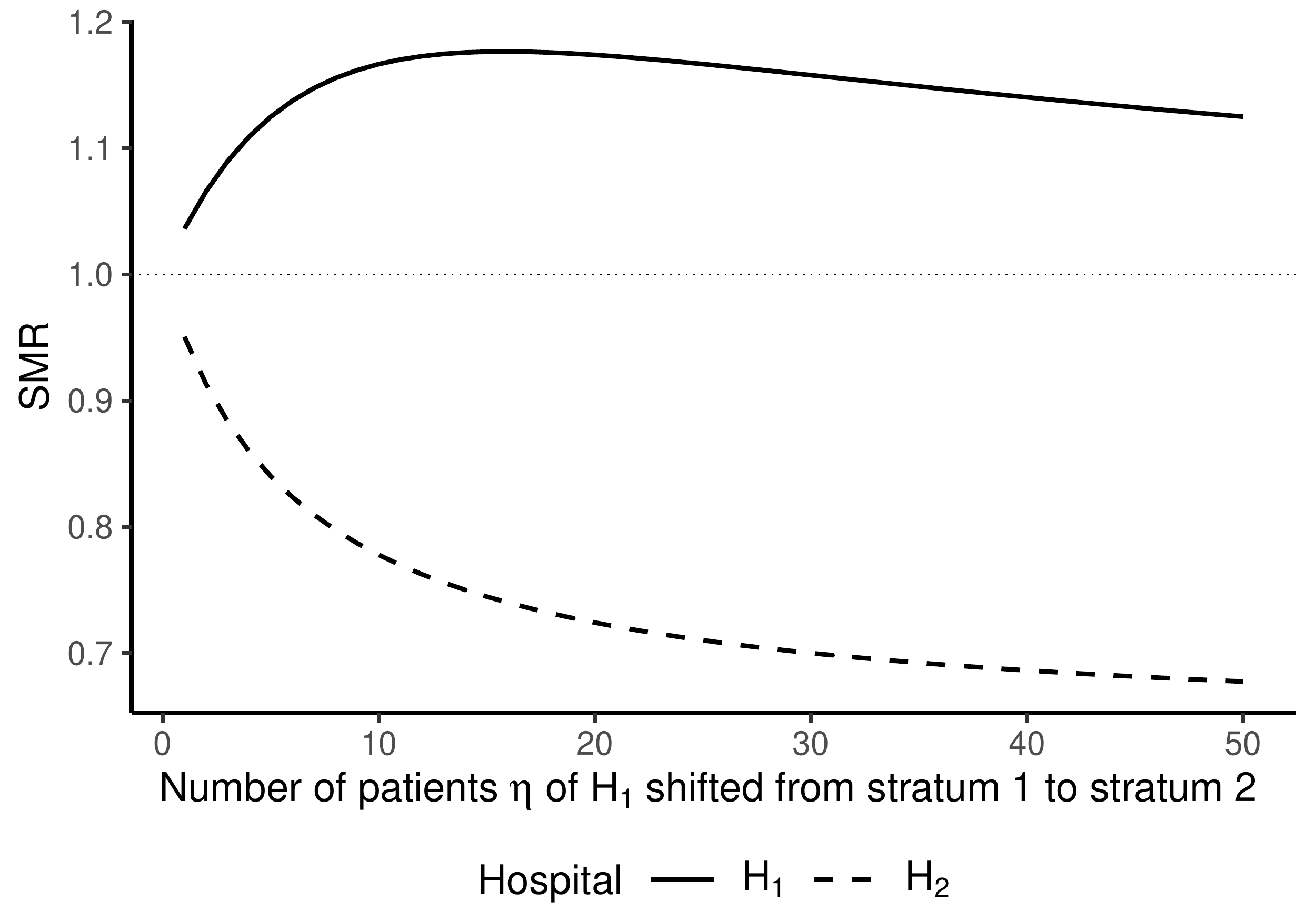}
	\caption{SMRs for different numbers of patients $\eta$ shifted from stratum 1 to stratum 2 in hospital~1  \label{fig:casemixint}}
\end{figure}

The SMRs of both hospitals for different values of $\eta$ are shown by Figure~\ref{fig:casemixint}. As the mortality rates of hospital 1 were higher or equal to those of hospital 2, the SMR of hospital 1 exceeds unity while the SMR of hospital 2 is below unity. If all patients of hospital 1 are allocated to stratum 1 ($\eta=0$), increases in $\eta$ lead to an increase in the SMR of hospital 1 and a decrease in the SMR of hospital 2. This behavior is in line with the fact that higher values of $\eta$ imply that more patients of hospital 1 are shifted from a stratum with a mortality rate equal to expected mortality to the stratum with a higher-than-expected mortality rate. However, at a certain number of patients allocated from stratum 1 to stratum 2, the SMR of hospital 1 does not change due to a change in $\eta$. At this point, the size of the hospital's SMR and its influence on the expected mortality rate of stratum 2 has become sufficiently large to meet the condition stated by Eq~\ref{eq:mechintmid}. When the number of patients treated in hospital 1 that is allocated from stratum 1 to stratum 2 is further increased, the SMR of hospital 1 even starts to decrease as implied by Eq~\ref{eq:mechintend}.

\begin{table}[h!]
	\centering
	\caption{Example of variations in case mix under internal standardization: parameter values}
	\begin{tabular}{c|cc|cc} \toprule
		Stratum & \multicolumn{2}{c|}{Hospital 1 ($H_1$)} & \multicolumn{2}{c|}{Hospital 2 ($H_2$)} \\ \midrule
		s     & $n_{1s}$ & $p_{1s}$ & $n_{2s}$ & $p_{2s}$ \\ \midrule
		1     & 50$-\eta$ & 0.1   & 25    & 0.1 \\
		2     & $\eta$ & 0.3   & 10    & 0.1 \\ \bottomrule
	\end{tabular}%
	\label{tab:casemixint}%
\end{table}%

\subsubsection*{Variations in hospital size under internal standardization}

In case of internal standardization, increasing the number of patients treated in hospital $h$ in all strata by factor $\lambda$ yields
\begin{equation}\label{eq:scaleint}
\text{SMR}_h^{\text{int}}(\lambda n_{h1},\dots,\lambda n_{hS}) =  \frac{\sum_{s=1}^{S} n_{hs} p_{hs} }{\sum_{s=1}^{S} n_{hs}p_s^{e,\text{int}} (\lambda n_{hs})} ,
\end{equation}
where
\begin{equation}\label{eq:lambps}
p_s^{e,\text{int}} (\lambda n_{hs}) = \frac{(\sum_{j=1}^{H} n_{js} p_{js}) + (\lambda-1) n_{hs} p_{hs}}{(\sum_{j=1}^{H} n_{js}) + (\lambda-1) n_{hs}} .
\end{equation}
Since Eq~\ref{eq:lambps} shows that the stratum-specific expected mortality rates depend on $\lambda$, the SMR does not fulfill the axiomatic requirement of scale insensitivity under internal standardization. For further investigation, the change in the SMR due to increasing the number of patients by factor $\lambda$ is defined as
\begin{align}
\Delta \text{SMR}_h^{\text{int}} &:= \text{SMR}_h^{\text{int}}(\lambda n_{h1},\dots,\lambda n_{hS}) - \text{SMR}_h^{\text{int}}(n_{h1},\dots,n_{hS}) \nonumber \\ 
&= \text{SMR}_h^{\text{int}}(n_{h1},\dots,n_{hS}) \cdot \left[ \frac{\bar{p}_h^{e,\text{int}}(n_{h1},\dots,n_{hS})}{\bar{p}_h^{e,\text{int}}(\lambda n_{h1},\dots, \lambda n_{hS})} - 1 \right] . \label{scaleint}
\end{align}
Eq~\ref{scaleint} implies that the SMR of hospital $h$ increases (decreases) due to an increase in hospital size if the hospital's expected overall mortality rate decreases (increases) due to scaling. Furthermore, the magnitude of change induced by scaling is (in absolute terms) higher (lower) for hospitals with higher (lower) initial SMRs. 

The condition determining the direction of change in the SMR may be expressed as
\begin{align}\label{casesmrint}
\Delta \text{SMR}_h^{\text{int}}  & > 0 \quad \text{if} \quad \sum_{s=1}^{S} \frac{n_{hs}}{n_h} [p_s^{e,\text{int}}(n_{hs}) - p_s^{e,\text{int}}(\lambda n_{hs})] > 0 , \\
\Delta \text{SMR}_h^{\text{int}}  & = 0 \quad \text{if} \quad \sum_{s=1}^{S} \frac{n_{hs}}{n_h} [p_s^{e,\text{int}}(n_{hs}) - p_s^{e,\text{int}}(\lambda n_{hs})] = 0 , \\
\Delta \text{SMR}_h^{\text{int}}  & < 0 \quad \text{if} \quad \sum_{s=1}^{S} \frac{n_{hs}}{n_h} [p_s^{e,\text{int}}(n_{hs}) - p_s^{e,\text{int}}(\lambda n_{hs})] < 0 .
\end{align}
Hence, the sign of Eq~\ref{scaleint} depends on the patient-share-weighted average of the differences $p_s^{e,\text{int}}(n_{hs}) - p_s^{e,\text{int}}(\lambda n_{hs})$ in stratum-specific expected mortality rates before and after scaling. Further analysis reveals that 
\begin{align}\label{eq:stratdiff}
p_s^{e,\text{int}}(n_{hs}) - p_s^{e,\text{int}}(\lambda n_{hs}) & > 0 \quad \text{if} \quad p_{hs} < \bar{p}_s(n_{hs}) ,  \\
p_s^{e,\text{int}}(n_{hs}) - p_s^{e,\text{int}}(\lambda n_{hs}) & = 0 \quad \text{if} \quad p_{hs} = \bar{p}_s(n_{hs}) , \\
p_s^{e,\text{int}}(n_{hs}) - p_s^{e,\text{int}}(\lambda n_{hs}) & < 0 \quad \text{if} \quad p_{hs} > \bar{p}_s(n_{hs}) . \label{eq:stratdiffend}
\end{align}
For a hospital with above-average stratum-specific mortality rates in all strata, Eqs~\ref{eq:stratdiff}-\ref{eq:stratdiffend} imply a decrease in the SMR when the scale of those hospital is increased. On the contrary, the SMR of a hospital performing better than average in all strata increases when its size is increased while holding case mix constant. In accordance with these results, it can be derived that
\begin{equation}\label{eq:smrintlimit}
\lim\limits_{\lambda \rightarrow \infty} \text{SMR}_h^{\text{int}}(\lambda n_{h1},\dots,\lambda n_{hS}) = \frac{\sum_{s=1}^{S} n_{hs} p_{hs} }{\sum_{s=1}^{S} n_{hs} \lim\limits_{\lambda \rightarrow \infty} p_s^{e,\text{int}} (\lambda n_{hs})} = 1 ,
\end{equation}
since Eq~\ref{eq:lambps} implies that $\lim\limits_{\lambda \rightarrow \infty} p_s^{e,\text{int}} (\lambda n_{hs}) = p_{hs}$. Hence, the SMR under internal standardization approaches (but does not cross) unity when the scale of a hospital is increased. These results reflects that the hospital is increasingly becoming its own reference when its size is increased because it increasingly dominates the value of the stratum-specific expected mortality rates.

For illustration of scale sensitivity under internal standardization, we considered three hospitals and three strata of patients (Table~\ref{tab:scaleint}). Hospitals 1 and 2 had patients in strata 1 and 2 and no patient belonging to stratum 3. Hospital 3 treated patients belonging to strata 2 and 3 but no patient belonging to stratum 1. In terms of mortality rates, hospital 2 performed better than the other hospitals in all strata. Hospital 2 performed better than hospital 3 in stratum 2. The patient numbers of hospital 1 in all strata were scaled by factor $\lambda$.

\begin{figure}[h!] \centering
	\includegraphics[width=.8\textwidth]{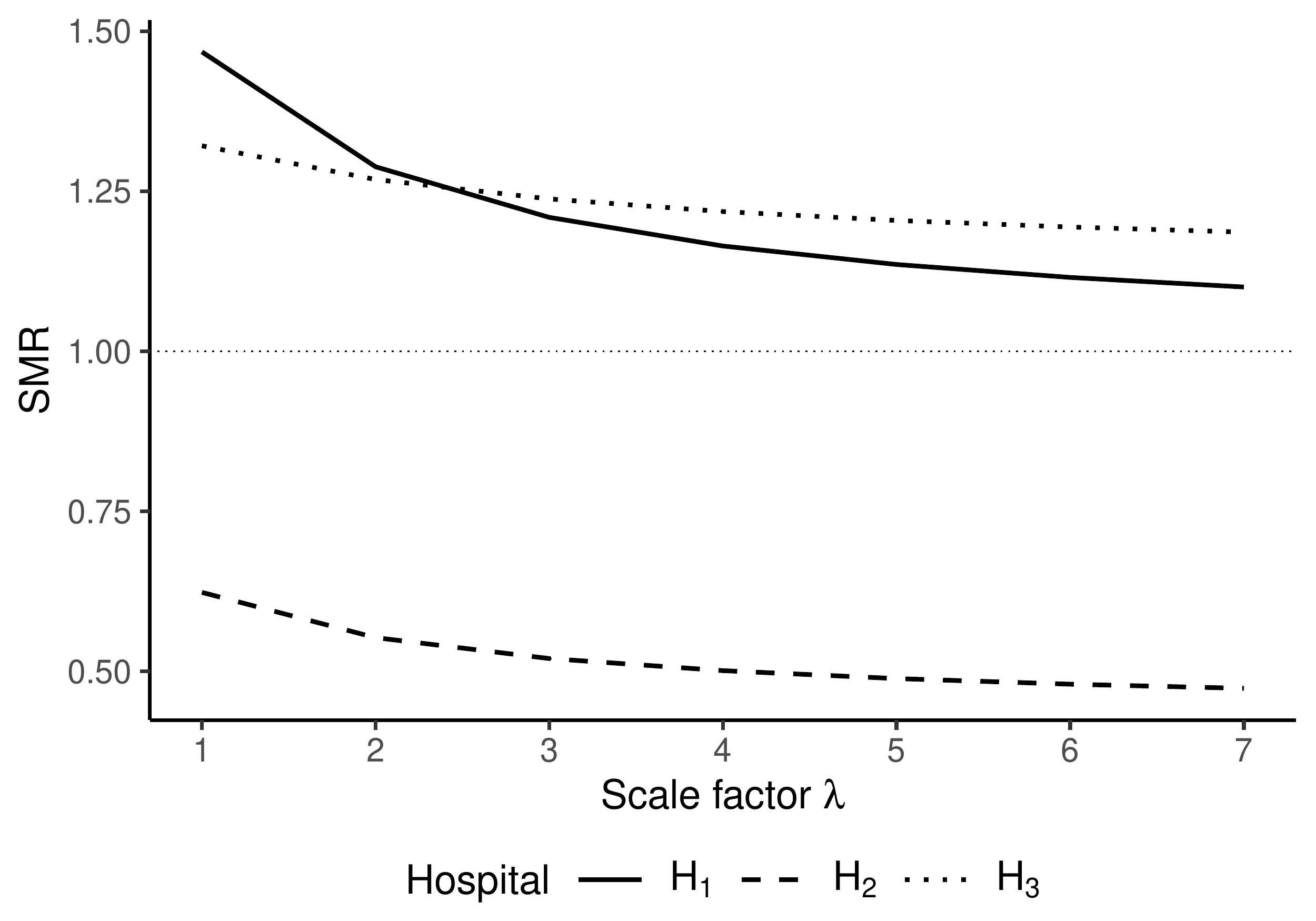}
	\caption{SMRs for different scale factors $\lambda$ affecting the size of hospital 1 \label{fig:scaleint}}
\end{figure}

As depicted by Fig~\ref{fig:scaleint}, in the initial situation ($\lambda = 1$) hospital 1 has the highest SMR while hospital 2 has the lowest SMR. The SMR of hospital 3 exceeds unity, indicating worse-than-average performance, but is lower than the SMR of hospital 1. Doubling the size of hospital 1 ($\lambda = 2$) leads to a decrease in the SMRs of all three hospitals. This is due to the increased weight of hospital 1 in the calculation of the expected mortality rates of strata 1 and 2. Since the stratum-specific mortality rates of hospital 1 are higher than the average mortality rates in the initial situation, this results in an increase in expected mortality rates (see Eqs~\ref{eq:stratdiff}-\ref{eq:stratdiffend}). However, the induced decrease in the SMR is strongest for hospital 1, implying that it becomes more close to hospital 3 in terms of the overall performance assessment. For further increased scales of hospital 1 ($\lambda \geq 3$), this trend continues and the SMR of hospital 1 becomes lower than the SMR of hospital 3.

\begin{table}[h!]
	\centering
	\caption{Example of variations of hospital size under internal standardization: parameter values}
	\begin{tabular}{c|cc|cc|cc} \toprule
		Stratum & \multicolumn{2}{c|}{Hospital 1 ($H_1$)} & \multicolumn{2}{c|}{Hospital 2 ($H_2$)} & \multicolumn{2}{c|}{Hospital 3 ($H_3$)} \\ \midrule
		s     & $n_{1s}$ & $p_{1s}$ & $n_{2s}$ & $p_{2s}$ & $n_{3s}$ & $p_{3s}$ \\ \midrule
		1     & $50\cdot \lambda$ & 0.3   & 50    & 0.1   & 0     & - \\
		2     & $50\cdot \lambda$ & 0.2   & 100   & 0.1   & 10    & 0.25 \\
		3     & 0     & -     & 0     & -     & 10    & 0.2 \\ \bottomrule
	\end{tabular}%
	\label{tab:scaleint}%
\end{table}%

\subsubsection*{Variations in actual mortality rates under internal standardization}

The marginal effect of an increase in the mortality rate of stratum $k$ in hospital $h$ can be derived as
\begin{align}\label{eq:me_int}
\text{ME}_{h,p_{hk}}^{\text{int}} &:= \frac{\partial \text{SMR}_h^{\text{int}}}{\partial p_{hk}} = \frac{n_{hk}}{n_h} \cdot \frac{1}{\bar{p}_h^{e,\text{int}}} - \text{SMR}_h^{\text{int}} \cdot \frac{n_{hk}}{n_h} \cdot \frac{1}{\bar{p}_h^{e,\text{int}}} \cdot \frac{\partial p_k^{e,\text{int}}}{\partial p_{hk}} \\
&= \frac{n_{hk}}{n_h} \cdot \frac{1}{\bar{p}_h^{e,\text{int}}} \cdot \left( 1 - \text{SMR}_h^{\text{int}} \cdot \frac{n_{hk}}{n_k} \right)
\end{align}
Note that the first term on the right-hand side of Eq~\ref{eq:me_int} is similar to the first term on the right hand side of Eq~\ref{eq:me_ext}. Thus, this term represents the direct effect of an increase in the stratum-specific mortality on the SMR of hospital $h$. However, when using an internal standard there is also an indirect effect, represented by the second term on the right hand side of Eq~\ref{eq:me_int}. This indirect effect emerges from the fact that the expected mortality rate $p_k^{e,\text{int}}$ depends on the the mortality rate $p_{hk}$ of patients included in this stratum treated in hospital $h$. Given that $\partial p_k^{e,\text{int}} / \partial p_{hk} = n_{hk}/n_k > 0$ if $n_{hk} > 0$, the second term on the right-hand side of Eq~\ref{eq:me_int} is negative, which indicates that the indirect effect counteracts the positive direct effect of an increase in the stratum-specific mortality rate. 

It follows that
\begin{align}\label{eq:condnegative}
\text{ME}_{h,p_{hk}}^{\text{int}} > 0 \quad \text{if} \quad \text{SMR}_h^\text{int} < \frac{n_k}{n_{hk}} , \\
\text{ME}_{h,p_{hk}}^{\text{int}} = 0 \quad \text{if} \quad \text{SMR}_h^\text{int} = \frac{n_k}{n_{hk}} , \\
\text{ME}_{h,p_{hk}}^{\text{int}} < 0 \quad \text{if} \quad \text{SMR}_h^\text{int} > \frac{n_k}{n_{hk}} . \label{eq:negme}
\end{align}
As shown by Eq~\ref{eq:negme}, the marginal effect of $p_{hk}$ may even be negative if the hospital's SMR is high and the hospital accounts for a large share of patients in stratum $k$, i.e. if $n_k/n_{hk}$ is small. This corresponds to the paradoxical situation in which increasing mortality in a stratum of patients treated in a specific hospital reduces the SMR of that hospital. Hence, the SMR under internal standardization does not fulfill the axiomatic requirement of strict monotonicity. Since $n_k/n_{hk} \geq 1$, Eq~\ref{eq:negme} further shows that a negative marginal effect can only arise in hospital with above-average SMRs, i.e.\ in hospitals with $\text{SMR}_h^{\text{int}} > 1$.

Considering a change in the mortality rates of all strata by the amount of $\mathrm{d}p_{hs} = \mathrm{d}p, s=1,\dots,S$ yields
\begin{equation}\label{eq:totaldiffint}
\mathrm{d} \text{SMR}_h^{\text{int}} = \sum_{s=1}^{S} \frac{\partial \text{SMR}_h^{\text{int}}}{\partial p_{hs}} \mathrm{d} p_{hs} = \frac{\mathrm{d} p}{\bar{p}_h^{e,\text{int}}} \left( 1 - \text{SMR}_h^{\text{int}} \sum_{s=1}^{S} \frac{n_{hs}}{n_h} \cdot \frac{n_{hs}}{n_k} \right).
\end{equation}
The expression describing the change in the internally standardized SMR (Eq~\ref{eq:totaldiffint}) differs from the expression for the change in the externally standardized SMR (Eq~\ref{eq:totaldiffext}) due to the factor in parentheses included in Eq~\ref{eq:totaldiffint}. This factor is smaller than 1 if $\text{SMR}_h^{\text{int}} > 0$, implying that the increase in the SMR of a hospital due to an increase in its overall mortality rate is, generally, smaller under internal than under external standardization. Moreover, the sign of Eq~\ref{eq:totaldiffint} is ambiguous since
\begin{align}\label{eq:dsmrdirection}
\mathrm{d} \text{SMR}_h^{\text{int}} > 0 \quad \text{if} \quad \text{SMR}_h^{\text{int}} < \frac{1}{\sum_{s=1}^{S} \frac{n_{hs}}{n_h} \cdot \frac{n_{hs}}{n_k}} , \\
\mathrm{d} \text{SMR}_h^{\text{int}} = 0 \quad \text{if} \quad \text{SMR}_h^{\text{int}} = \frac{1}{\sum_{s=1}^{S} \frac{n_{hs}}{n_h} \cdot \frac{n_{hs}}{n_k}} , \\
\mathrm{d} \text{SMR}_h^{\text{int}} < 0 \quad \text{if} \quad \text{SMR}_h^{\text{int}} > \frac{1}{\sum_{s=1}^{S} \frac{n_{hs}}{n_h} \cdot \frac{n_{hs}}{n_k}} .
\end{align}
An increase in the overall mortality rate of a hospital therefore may reduce its SMR if the hospital's SMR exceeds the threshold $(\sum_{s=1}^{S} \frac{n_{hs}}{n_h} \cdot \frac{n_{hs}}{n_k})^{-1} \geq 1$. This threshold takes on the value of 1 if all patients of the hospital are concentrated in a specific stratum $k$ ($n_{hk}/n_h = 1$) and the hospital accounts for all patients belonging to this stratum ($n_{hk}/n_k = 1$). Thus, paradoxical effects of mortality rate increases on the SMR may particularly arise in specialized hospitals treating specific patient groups that are seldom treated in other hospitals.

For illustration, we considered three hospitals and two strata of patients (Table~\ref{tab:actmortint}). Hospitals 1 and 2 treated patients belonging to strata 1 and 2 whereas hospital 3 treated patients belonging to stratum 2 only. With respect to stratum 2, hospital 1 had the highest mortality rate whereas hospital 3 had the lowest mortality rate. In the following, the mortality rate $p_{11} \in [0,1]$ in stratum 1 of hospital 1 is varied for different shares $w_{11} \in \{0.6,0.8,1\}$ of patients in stratum 1 treated in hospital 1. The larger $w_{11}$, the higher the share of patients belonging to stratum 1 that were treated in hospital 1. 

\begin{figure}[h!] \centering
\includegraphics[width=.8\textwidth]{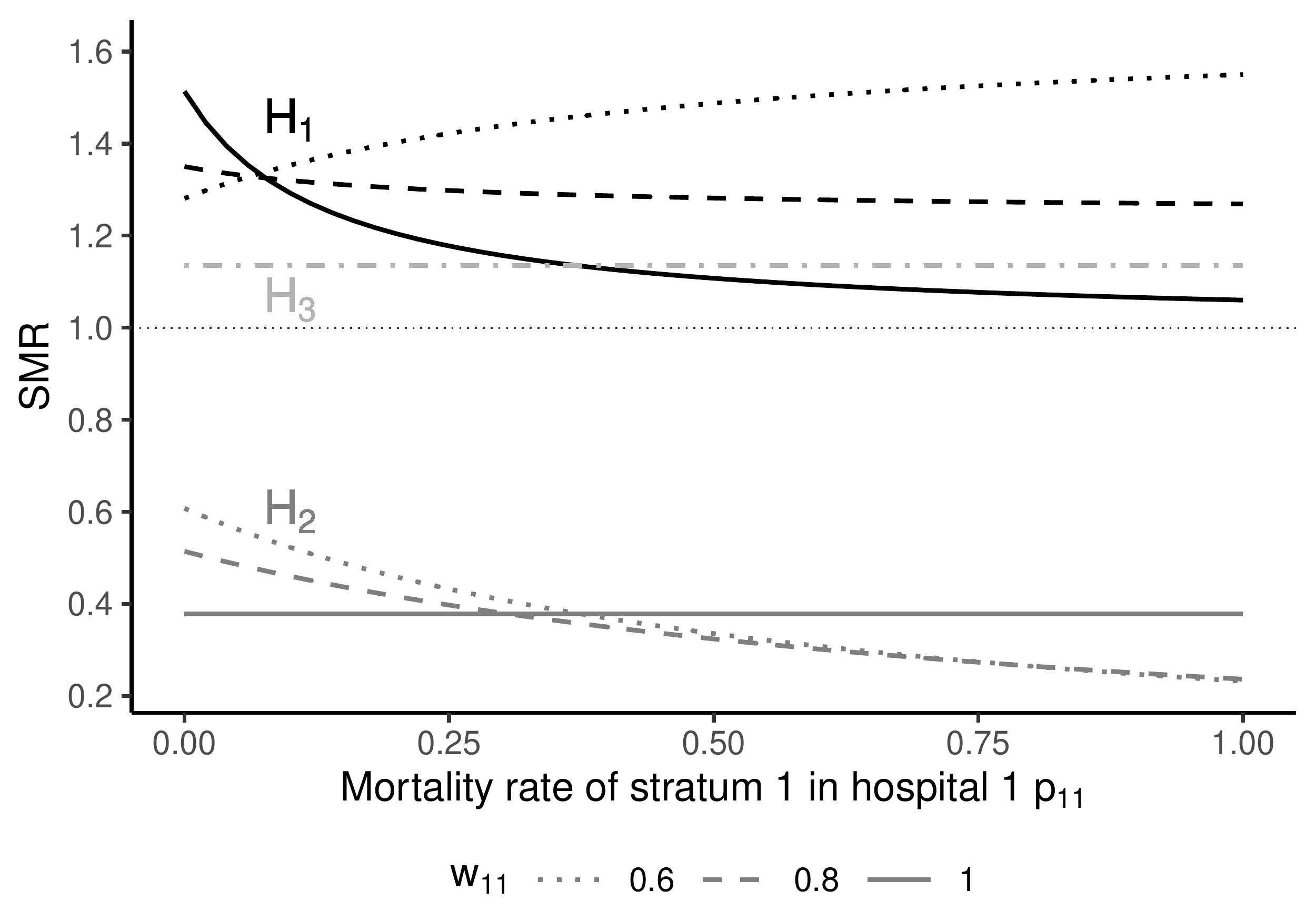}
\caption{SMRs for different mortality rates $p_{11}$ and patient shares $w_{11}$ of hospital 1 \label{fig:hypsmr}}
\end{figure}

The results are shown in Fig~\ref{fig:hypsmr}. If 60\% of all patients in stratum 1 are allocated to hospital~1 ($w_{11}=0.6$), an increase in the mortality rate of these patients is associated with an increase in the SMR of hospital 1 and a decrease of the SMR of hospital 2. Note that the SMR of hospital 3 is not affected by variations in the mortality rate of stratum 1 as no patients belonging to this stratum were treated in hospital 3. When the share of patients belonging to stratum 1 allocated to hospital 1 is increased to 80\% ($w_{11}=0.8$), the SMR of hospital 1 decreases in the morality rate $p_{11}$. This illustrates the paradoxical situation captured by Eq~\ref{eq:negme}, in which increasing mortality in a stratum of patients reduces the SMR of the considered hospital. In the extreme case in which all patients in stratum 1 are treated in hospital 1 ($w_{11}=1$), the inverse relationship between stratum-specific mortality and SMR of hospital 1 gets even more pronounced. If the mortality rate of stratum 1 in hospital 1 reaches 100\%, the SMR of hospital 1 gets close to unity. Note that this scenario also illustrates that the SMR of hospital 1 can become lower than the SMR of hospital 3 although the mortality rate in stratum 2 (the only stratum with a positive number of patients treated in hospital 3) is 10\% lower in hospital 3 than in hospital 1.

\begin{table}[h!]
	\centering
	\caption{Example of variations in actual mortality rates under internal standardization: parameter values}
	\begin{tabular}{c|cc|cc|cc} \toprule
		Stratum & \multicolumn{2}{c|}{Hospital 1 ($H_1$)} & \multicolumn{2}{c|}{Hospital 2 ($H_2$)} & \multicolumn{2}{c|}{Hospital 3 ($H_3$)} \\ \midrule
		s     & $n_{1s}$ & $p_{1s}$ & $n_{2s}$ & $p_{2s}$ & $n_{3s}$ & $p_{3s}$ \\ \midrule
		1     & 100$\cdot w_{11}$ & $p_{11}$ & 100$\cdot (1-w_{11})$ & 0.1   & 0     & - \\
		2     & 50    & 0.4   & 50    & 0.1   & 40    & 0.3 \\ \bottomrule
	\end{tabular}%
	\label{tab:actmortint}%
\end{table}%

\subsubsection*{Variations in expected mortality rates under internal standardization}
Analogous to the SMR under external standardization, the SMR under internal standardization is affected by changes in expected mortality rates, which leads to a violation of the dominance principle. However, due to the endogeneity of the stratum-specific expected mortality rates $p_s^{e,\text{int}}$ under internal standardization, variations in these expected mortality rates may be driven by variations in the mortality rates and patient compositions of all hospitals in the sample. The analyses shown above already highlighted effects of variations in mortality rates and patient composition of a hospital on its own SMR. In the following, we examine the influence of other hospitals.

First, the change in the SMR due to a change in the expected mortality rate of stratum $k$ is expressed as
\begin{equation}\label{eq:expmortint}
\mathrm{d} \text{SMR}_h^{\text{int}} = - \text{SMR}_h^{\text{int}} \cdot \frac{n_{hk}}{n_k} \cdot \frac{1}{\bar{p}_h^{e,\text{int}}} \cdot \mathrm{d} p_k^{e,\text{int}} .
\end{equation}
Eq~\ref{eq:expmortint} indicates that the SMR decreases (increases) if the expected mortality rate increases (decreases). Using Eq~\ref{eq:avmort}, the marginal effect of an increase in the mortality rate of stratum $k$ in hospital $i \neq h$ is
\begin{equation}\label{eq:peintpi}
\frac{\partial p_k^{e,\text{int}}}{\partial p_{ik}} = \frac{n_{ik}}{n_k} > 0 \quad \text{if} \quad n_{ik} > 0.
\end{equation}
In combination, Eqs~\ref{eq:expmortint}-\ref{eq:peintpi} imply that the SMR of a hospital decreases when the stratum-specific mortality rates of other hospitals in the sample are increased. The reason is that such increases in mortality rates unambiguously increase the expected mortality rates of the affected strata. 

This effect is illustrated by the behavior of the SMR of hospital 2 in Fig~\ref{fig:hypsmr}, which decreases in the mortality rate of stratum 1 in hospital 1 as long as hospital 2 accounts for a positive number of patients in that stratum.

Second, adding $\eta$ patients to stratum $k$ in hospital $i$ implies
\begin{equation}\label{eq:expnhl}
p_k^{e,\text{int}}(n_{ik}+\eta) - p_k^{e,\text{int}}(n_{ik}) = \frac{p_{ik} - \bar{p}_k(n_{ik})}{1 + \eta^{-1}n_k} .
\end{equation}
Hence, increasing the size of stratum $k$ in hospital $i$ (and, thus, its share in the total number of patients belonging to stratum $k$) increases (decreases) the expected mortality rate of that stratum if hospital $i$'s mortality rate of that stratum $p_{ik}$ is higher (lower) than the average mortality rate of that stratum $\bar{p}_k(n_{ik})$. The direction of change in the SMR of a hospital $h$ (Eq~\ref{eq:expmortint}) due to a change in other hospitals' stratum-specific patient numbers therefore depends on whether those hospitals perform better or worse than average in the affected strata. 

An illustration of this result is given by the variation in the SMR of hospital 2 in Fig~\ref{fig:casemixint}. Since hospital 1 performs worse than average in stratum 2, increasing the number of patients $\eta$ in this hospital belonging to that stratum reduces the SMR of hospital 2.

\subsection*{Summary of results}
Evaluating the derived properties of the SMR using the five axiomatic requirements formulated above yielded differences between external and internal standardization (Table~\ref{tab:summary}). Under external standardization, the SMR fulfills the requirements of strict monotonicity and scale insensitivity but violates the requirement of case-mix insensitivity, the equivalence principle, and the dominance principle. All axiomatic requirements not fulfilled by the SMR under external standardization are also not fulfilled by the SMR under internal standardization due to similarity in their mathematical structure. Additionally, higher mortality rates may induce lower SMR values and the SMR of large hospitals is driven towards unity under internal standardization. The internally standardized SMR therefore also violates the requirements of strict monotonicity and scale insensitivity and, thus, fulfills none of the postulated axiomatic requirements.

\begin{table}[h!] \caption{Fulfillment of axiomatic  requirements by standardization approach \label{tab:summary}} \centering
\begin{tabular}{lcc} \toprule
 & SMR under  & SMR under \\
Axiomatic requirement & external standardization & internal standardization \\ \midrule
Strict monotonicity & yes & no \\
Case-mix insensitivity & no & no \\
Scale insensitivity & yes & no \\
Equivalence principle & no & no \\
Dominance principle & no & no \\ \bottomrule
\end{tabular}
\end{table}

\section*{Discussion}
This paper proposed five axiomatic requirements for risk standardized mortality measures (strict monotonicity, case-mix insensitivity, scale insensitivity, equivalence principle, dominance principle). Given these axiomatic requirements, properties of the SMR were formally investigated and evaluated.

The results of our analyses indicate that several properties of the SMR hamper valid assessment and comparison of hospital performance based on this measure. This finding has very high public health relevance, as clinicians, healthcare decision makers, the public, and all users of quality of care information based on SMRs are confronted with potentially biased information and, thus, may draw inappropriate conclusions. Effects of variations in case mix on the SMR were found to depend not only on hospital size and the initial patient composition of a hospital but also on the size of its SMR. Variations in actual mortality rates depend on the hospital's expected overall mortality rate and, thus, on its case mix. Under external standardization, the stratum-specific \textit{expected} mortality rates have crucial influence on the size of the SMR. Paradoxically, variations in these expected mortality rates may reverse the rank of two hospitals although one of the hospitals unambiguously performs better than the other in terms of \textit{actual} mortality rates. 

While hospital size has no effect on the SMR under external standardization, this desirable property of scale insensitivity is absent under internal standardization. In this case, the SMR of large hospitals is, ceteris paribus, more close to 1 than the hospital of small hospitals. This results is driven by the fact that large hospitals have more influence on expected mortality rates than small hospitals under internal standardization. This influence on expected mortality rates also modifies the effect of variations in actual mortality rates on the SMR. In extreme cases, higher actual mortality rates may be related to a lower SMR of the considered hospital. This paradoxical effect particularly may arise in specialized hospitals that almost exclusively treated specific patient groups.

In summary, our findings significantly extend previous research on properties of the SMR \cite{Manktelow2014, Glance1999, Glance2000, Kahn2007, Pouw2013} by formally deriving expressions and conditions describing the behavior of the SMR. In this way, this study provides a comprehensive and exact characterization of this commonly used hospital performance measure.

\subsection*{Limitations and prospects}
The analyses presented in this paper provide a clear description of central properties of the SMR. However, although we constructed hypothetical examples illustrating these properties, we did not provide empirical examples based on real-world data. Presumably, the extent to which the described drawbacks of the SMR are empirically relevant depends on the considered indication, the choice of risk factors used to define patient strata, and the similarity of the considered hospitals with respect to case mix and mortality rates. While investigating these issues in specific settings is beyond the scope of this paper, future studies may examine the insights highlighted in this paper empirically.

Furthermore, while this study revealed several undesirable properties of the SMR under both external and internal standardization, it did not provide an alternative measure of hospital performance. Some studies point to certain advantages of measures like the comparative mortality figure (CMF) or excess risk (ER) \cite{Roessler2019,Varewyck2014}. However, there is a lack of formal analysis and comparison in the literature. Systematically analyzing and developing suitable hospital performance measures therefore may be a promising route for further research. 

Finally, there may be relevant properties of the SMR that were not investigated above. Formal examination of those properties may be a valuable complement to the analyses presented in this paper.

\subsection*{Practical implications}
Contrary to internal standardization, external standardization ensures strict monotonicity and scale insensitivity. Hence, external standardization should generally be preferred over internal standardization in practical applications. This is particularly true when the number of analyzed hospitals is small or when there are large and/or specialized hospitals that almost exclusively treated specific patient groups. Nonetheless, practitioners should be aware of the potential drawbacks related to the use of the SMR under both standardization approaches. The SMR generally violates the requirement of case-mix insensitivity, the equivalence principle, and the dominance principle. Particularly in the presence of large heterogeneity of the analyzed hospitals in terms of case mix and mortality rates, the SMR cannot be trusted. As a general recommendation, empirical studies therefore should assess and report the degree of heterogeneity of the considered hospitals and take effects of heterogeneity into account when interpreting calculated SMRs.

\end{document}